\documentclass{SciPost}

\hypersetup{
    colorlinks,
    linkcolor={red!50!black},
    citecolor={blue!50!black},
    urlcolor={blue!80!black}
}

\usepackage[bitstream-charter]{mathdesign}
\usepackage{booktabs}
\usepackage{array}
\DeclareSymbolFont{usualmathcal}{OMS}{cmsy}{m}{n}
\DeclareSymbolFontAlphabet{\mathcal}{usualmathcal}

\fancypagestyle{SPstyle}{
\fancyhf{}
\lhead{\colorbox{scipostblue}{\bf \color{white} ~SciPost Physics Codebases }}
\rhead{{\bf \color{scipostdeepblue} ~Submission }}

\fancyfoot[C]{\textbf{\thepage}}
}

\begin{document}

\pagestyle{SPstyle}

\begin{center}{\Large \textbf{\color{scipostdeepblue}{
LaQX: A JAX Toolkit for Neural Quantum State Simulations of Lattice Quantum Many-Body Models
}}}\end{center}

\begin{center}\textbf{
Yuntian Gu\textsuperscript{1,2$\star$},
Chaohui Fan\textsuperscript{1,3},
Ruichen Li\textsuperscript{1},
Haoxiang Chen\textsuperscript{1,4},\\
Weizhong Fu\textsuperscript{1,4},
Yixiao Chen\textsuperscript{1},
Dingshun Lv\textsuperscript{1},
Weiluo Ren\textsuperscript{1},\\
Tao Xiang\textsuperscript{3$\dagger$},
Liwei Wang\textsuperscript{2$\ddagger$},
Mingpu Qin\textsuperscript{5$\S$},
Xiang Li\textsuperscript{1$\P$}
}\end{center}

\begin{center}
{\bf 1} ByteDance Seed, Beijing 100098, China
\\
{\bf 2} School of Intelligence Science and Technology, Peking University, Beijing 100871, China
\\
{\bf 3} School of Physical Sciences, University of Chinese Academy of Sciences, Beijing 100049, China
\\
{\bf 4} School of Physics, Peking University, Beijing 100871, China
\\
{\bf 5} School of Physics and Astronomy, Shanghai Jiao Tong University, Shanghai 200240, China
\\[\baselineskip]
$\star$ \href{mailto:guyuntian@bytedance.com}{\small guyuntian@bytedance.com}\,,\quad
$\dagger$ \href{mailto:txiang@iphy.ac.cn}{\small txiang@iphy.ac.cn}
\\
$\ddagger$ \href{mailto:wanglw@pku.edu.cn}{\small wanglw@pku.edu.cn}\,,\quad
$\S$ \href{mailto:qinmingpu@sjtu.edu.cn}{\small qinmingpu@sjtu.edu.cn}
\\
$\P$ \href{mailto:lixiang.62770689@bytedance.com}{\small lixiang.62770689@bytedance.com}\,\quad
\end{center}

\section*{\color{scipostdeepblue}{Abstract}}
\textbf{\boldmath{%
Neural quantum states provide accurate variational representations for interacting lattice fermions and spins. Turning a new ansatz into a reproducible calculation, however, requires coordinated implementations of the Hamiltonian, sampling, optimization, observables, and accelerator parallelism. We present LaQX (Lattice Quantum simulation based on jaX), an open-source neural quantum state library based on JAX that organizes these components through common functional interfaces. Beyond ground-state variational Monte Carlo, LaQX supports excited-state calculations, Green's Function Monte Carlo refinement, adiabatic evolution, and real-time evolution with time-dependent variational Monte Carlo. It consolidates methods developed in recent high-accuracy neural quantum state studies of the Hubbard model and frustrated magnets \cite{gu_hubbard_2025,gu_pareto_2026,fan_disentangling_2026}, and the package has been used to investigate the triangular-lattice Hofstadter-Hubbard model \cite{gu_spincharge_2026}. We document these capabilities with reproducible calculations for Hubbard, $J_1$-$J_2$ Heisenberg, altermagnetic Hubbard, and interacting Hofstadter models. Function-oriented modules separate the scientific components and limit the source context needed for extensions by human developers and large language models. Source code, documentation, and executable examples are available at \url{https://github.com/bytedance/laqx}.
}}

\vspace{\baselineskip}

%%%%%%%%%% BLOCK: Copyright information
% This block will be filled during the proof stage, and finilized just before publication.
% It exists here only as a placeholder, and should not be modified by authors.
\noindent\textcolor{white!90!black}{%
\fbox{\parbox{0.975\linewidth}{%
\textcolor{white!40!black}{\begin{tabular}{lr}%
  \begin{minipage}{0.6\textwidth}%
    {\small Copyright attribution to authors. \newline
    This work is a submission to SciPost Physics Codebases. \newline
    License information to appear upon publication. \newline
    Publication information to appear upon publication.}
  \end{minipage} & \begin{minipage}{0.4\textwidth}
    {\small Received Date \newline Accepted Date \newline Published Date}%
  \end{minipage}
\end{tabular}}
}}
}
%%%%%%%%%% BLOCK: Copyright information

% For convenience during refereeing we turn on line numbers:
% \linenumbers

\vspace{10pt}
\noindent\rule{\textwidth}{1pt}
\tableofcontents
\noindent\rule{\textwidth}{1pt}
\vspace{10pt}

\section{Introduction}
\label{sec:intro}
Lattice models connect microscopic Hamiltonians to collective quantum phenomena. Their numerical solution is nevertheless difficult because the Hilbert-space dimension grows exponentially with system size. Exact diagonalization (ED) is consequently restricted to finite clusters \cite{sandvik2010computational}, density-matrix renormalization group \cite{white1992dmrg} and matrix product state (MPS) methods \cite{schollwock2011dmrg} are most effective in one dimension and quasi-one-dimensional geometries, and higher-dimensional tensor-network contractions require additional approximations \cite{cirac2021tensor,xiang2023density}. Quantum Monte Carlo \cite{foulkes2001qmc,sandvik2010computational,assaadWorldlineDeterminantalQuantum2008,Becca_Sorella_2017,AFQMC-lecture-notes-2013} can reach much larger systems when a non-negative sampling weight is available, but frustration and fermionic statistics can produce a sign problem whose generic resolution is computationally hard \cite{lohSignProblemNumerical1990,troyer2005sign}. Comparisons on paradigmatic models such as the Hubbard model \cite{hubbard1963} therefore continue to rely on complementary numerical methods with different controlled limits \cite{leblanc2015hubbard,qinHubbardModelComputational2022,arovasHubbardModel2022}.

Neural quantum states (NQS) add a flexible class of variational wavefunctions to this landscape. In variational Monte Carlo (VMC) \cite{foulkes2001qmc,Becca_Sorella_2017}, a neural network represents the many-body amplitude, and its parameters are optimized using stochastic estimates over the Born distribution \cite{carleo2017solving}. The field now encompasses a broad range of architectures and applications \cite{lange2024review,medvidovic2024review}. Early work combined restricted Boltzmann machines with conventional fermionic pair-product states \cite{nomura2017rbm}, while convolutional networks addressed frustrated magnets \cite{choo2019j1j2} and autoregressive recurrent networks enabled direct sampling \cite{hibatallah2020rnn}. Subsequent developments introduced symmetry-aware deep networks for frustrated spin systems \cite{roth2023frustrated}, neural backflow and hidden-fermion constructions for lattice fermions \cite{luo2019backflow,moreno2022hidden,liu2024unifying}, and variational protocols for excited states and dynamics \cite{nes_science,schmitt2020dynamics}.

Recent studies have used hidden-fermion Pfaffian states \cite{roth2025superconductivity}, Transformer backflow \cite{gu_hubbard_2025,viteritti2026bias}, convolutional backflow \cite{gu_pareto_2026}, and symmetry-preserving pairing states \cite{rende2026superconductivity} to investigate the ground states of the square-lattice $t$-$t'$ Hubbard model. Large-scale NQS calculations have also examined topological superconductivity and quantum criticality in the triangular-lattice Hofstadter-Hubbard model \cite{gu_spincharge_2026,roth2026hofstadter_superconductivity}. Combining neural networks with tensor networks provides another route to correlated variational states \cite{fan_disentangling_2026}. These studies compare energies and order-specific correlations across variational families to distinguish competing phases. The range of fermionic, spin, excited-state, and dynamical applications motivates software that can accommodate new methods without duplicating the full VMC workflow.

Reusable software has developed alongside these methods. NetKet and jVMC provide general NQS and VMC infrastructure \cite{vicentini2022netket,schmitt2022jvmc}, while Quantax is a JAX package for flexible lattice many-body and NQS calculations \cite{quantax_software}. In the complementary setting of first-quantized, real-space electronic structure, FermiNet established the accuracy of antisymmetric neural wavefunctions \cite{pfau2020abinitio,spencer2020better}, and JaQMC provides an open JAX framework for neural-network VMC and diffusion Monte Carlo \cite{ren2023jaqmc,jaqmc_software}. Together with recent GPU-vectorized tensor-network VMC \cite{du_vectorized_2026}, these efforts show that accelerator performance depends not only on the cost of a single wavefunction evaluation, but also on batched sampling, differentiation, and linear algebra. They also establish the value of independent codebases with different physical scope and extension patterns.

Built on JAX \cite{jax2018}, LaQX targets interacting lattice fermions and spins. It provides neural backflow and tensor-network states, symmetry projection, Metropolis-Hastings sampling, first-order and natural-gradient optimization, and observable evaluation. The same interfaces support ground-state VMC with optional Green's Function Monte Carlo (GFMC) \cite{trivedi1990gfmc,sorella1998gfmc} refinement, joint optimization of low-lying states with Natural Excited States (NES) \cite{nes_science}, measurements on trained states, adiabatic evolution by variational continuation, and real-time evolution with time-dependent VMC (tVMC).

LaQX consolidates the computational framework developed for neural quantum state calculations of the doped Hubbard model, including Transformer-backflow states and the MARCH optimizer \cite{gu_hubbard_2025}, followed by the SCALE and ACE convolutional backflow architectures \cite{gu_pareto_2026}. It also incorporates architecture that combine convolutional neural networks (CNNs) with MPS, denoted CNN-MPS, for frustrated magnets \cite{fan_disentangling_2026}. The convolutional-backflow and CNN-MPS studies report variational energies among the best available for their respective fermionic and spin benchmarks. LaQX has subsequently been applied to the triangular-lattice Hofstadter-Hubbard model, where NQS and independent projected entangled pair state (PEPS) calculations provided complementary real-space evidence of spin-charge separation \cite{gu_spincharge_2026}. The present release brings these developments together in a documented package that supports multiple ansatz families and physical workflows.

The implementation is deliberately function-oriented. Wavefunctions, physical operators, samplers, and parameter updates are connected through small interfaces, with numerical state passed explicitly between them. This separation makes JAX transformations directly applicable and allows many scientific extensions to remain local to one component. It also supports programming with large language models (LLMs), including the conversational workflow known as \textit{vibe coding} \cite{sarkar2025vibe}. Limiting cross-component dependencies reduces the source context needed for an extension. Section~\ref{sec:vibe-coding} illustrates these extension paths.

The remainder of this paper is organized as follows. Section~\ref{sec:algorithm} defines the numerical methods and notation. Section~\ref{sec:software} describes the software architecture. Section~\ref{sec:examples} presents calculations for Hubbard, $J_1$-$J_2$ Heisenberg, altermagnetic, and interacting Hofstadter models.

\section{Algorithms}
\label{sec:algorithm}

\subsection{Variational Monte Carlo}
\label{sec:vmc}

Let $\mathbf n=(n_1,\ldots,n_L)$ contain the $L$ local occupation variables of a fermionic or spin configuration, and let $\psi_\theta(\mathbf n)$ be its amplitude under a variational state with parameters $\theta$. The variational total energy is
\begin{equation}
E(\theta)
=
\frac{\langle \psi_\theta | H | \psi_\theta\rangle}
{\langle \psi_\theta | \psi_\theta\rangle}.
\end{equation}
Introducing the normalized Born distribution
\begin{equation}
p_\theta(\mathbf n)
=
\frac{|\psi_\theta(\mathbf n)|^2}{\sum_{\mathbf n'}|\psi_\theta(\mathbf n')|^2}
\end{equation}
with $\psi_\theta(\mathbf n)=\langle \mathbf n | \psi_{\theta} \rangle$
gives $E(\theta)=\langle E_{\mathrm{loc}}\rangle_{p_\theta}$, where
\begin{equation}
E_{\mathrm{loc}}(\mathbf n)
=
\frac{\langle \mathbf n | H| \psi_{\theta} \rangle}{\psi_\theta(\mathbf n)}.
\label{eq:local-energy}
\end{equation}
For real variational coordinates $\theta_a$ (including separate real and imaginary coordinates when needed), define the logarithmic derivative $O_a(\mathbf n)=\partial_{\theta_a}\log\psi_\theta(\mathbf n)$. For a Hermitian Hamiltonian independent of $\theta$, differentiating the total energy $E(\theta)$ gives
\begin{equation}
\partial_{\theta_a}E
=
2\,\mathrm{Re}\left[
\left\langle O_a^* E_{\mathrm{loc}}\right\rangle
-
\left\langle O_a^*\right\rangle
\left\langle E_{\mathrm{loc}}\right\rangle
\right].
\label{eq:energy-gradient}
\end{equation}

\subsection{Neural wavefunction architectures}
\label{sec:ansatz}

SCALE and ACE are convolutional neural backflow states for lattice fermions \cite{gu_pareto_2026}, while CNN-MPS combines convolutional features with a tensor-network output for spins \cite{fan_disentangling_2026}. Their feature extractors use the same elementary operations: occupation embeddings, spatial convolutions, and site-wise multilayer perceptrons (MLPs) \cite{rumelhart1986learning}. We define these operations first, then specify how each architecture constructs the many-body amplitude.

\paragraph{Occupation embeddings and convolutional features.}
Let $N_s=L_xL_y$ denote the number of spatial sites, and let $q_i$ label the local basis state at site $i$. For spinful fermions, $q_i=n_{i\uparrow}+2n_{i\downarrow}$ distinguishes the four states $|0\rangle$, $|\uparrow\rangle$, $|\downarrow\rangle$, and $|\uparrow\downarrow\rangle$; for a spin-$1/2$ or spin-polarized occupation basis, $q_i\in\{0,1\}$. A trainable, site-dependent lookup table $\mathbf E$ maps this discrete label to a feature vector with $d_h$ channels,
\begin{equation}
\mathbf h_i^{(0)}=\mathbf E_{i,q_i}\in\mathbb R^{d_h}.
\label{eq:embedding}
\end{equation}
A convolution then mixes features at nearby sites. For a $k\times k$ kernel with odd $k$, define $\mathcal N_k=\{-(k-1)/2,\ldots,(k-1)/2\}^2$ and
\begin{equation}
[\mathcal C^{(\ell)}(\mathbf h)]_{i\beta}
=b_\beta^{(\ell)}+
\sum_{\boldsymbol\delta\in\mathcal N_k}\sum_{\alpha=1}^{d_h}
K_{\boldsymbol\delta\alpha\beta}^{(\ell)}h_{i+\boldsymbol\delta,\alpha}.
\label{eq:convolution}
\end{equation}
Here $\ell$ labels the layer, $\alpha$ and $\beta$ label input and output channels, and $K^{(\ell)}$ and $b^{(\ell)}$ are the trainable kernel and bias shared across sites. Indices wrap in a periodic direction, while features outside an open boundary are set to zero. The examples use the element-wise sigmoid linear unit (SiLU) activation \cite{elfwing2018sigmoid}, $\sigma(x)=x/(1+e^{-x})$. A residual block adds the transformed features to its input, $\mathbf h_i+\sigma[\mathcal C(\mathbf h)_i]$. Although convolutional weights are shared, the site-dependent embeddings do not by themselves impose translation symmetry on the state.

After the spatial blocks, an MLP acts separately on each site's features:
\begin{equation}
\begin{aligned}
\mathbf u_i^{(0)}&=\mathbf h_i^{(L_{\mathrm C})},\qquad \mathbf g_i=\mathbf u_i^{(L_{\mathrm M})},\\
\mathbf u_i^{(m+1)}&=\sigma\!\left(W^{(m)}\mathbf u_i^{(m)}+\mathbf b^{(m)}\right).
\end{aligned}
\label{eq:site-mlp}
\end{equation}
$L_{\mathrm C}$ and $L_{\mathrm M}$ count the convolutional and MLP layers. The MLP widths may differ from $d_h$. Its weights are shared across sites, but it introduces no additional spatial mixing: the receptive field, namely the input sites on which $\mathbf g_i$ depends, is set by the convolutional stack.

\paragraph{Determinantal backflow.}
For $N$ fermions, an affine output map produces an orbital table $M_\theta(\mathbf n)$ with $N$ columns and $n_sN_s$ rows, where $n_s=2$ for spinful and $n_s=1$ for spin-polarized fermions:
\begin{equation}
M_{(i,s),j}(\mathbf n)=\sum_\mu W^{\mathrm{out}}_{sj\mu}g_{i\mu}(\mathbf n)+b^{\mathrm{out}}_{sj}.
\label{eq:orbital-map}
\end{equation}
The indices $s$, $j=1,\ldots,N$, and $\mu=1,\ldots,d_g$ label the spin channel, orbital, and output feature of the MLP, respectively; $d_g$ is its output width. If $I_a(\mathbf n)$ is the $a$th occupied spin orbital, the single-determinant setting used here is
\begin{equation}
\Phi_{aj}(\mathbf n)=M_{I_a(\mathbf n),j}(\mathbf n),\qquad
\psi_\theta(\mathbf n)=\det\Phi_\theta(\mathbf n).
\label{eq:backflow}
\end{equation}
Thus an $N\times N$ matrix is obtained by selecting occupied rows from the orbital table. Unlike a fixed-orbital Slater determinant, each row depends on the surrounding occupations through the neural features, introducing backflow correlations \cite{luo2019backflow}. ACE can produce complex orbitals by combining separate real and imaginary affine outputs. When requested, lattice-symmetry projection forms a character-weighted sum of transformed amplitudes, including the fermionic reordering signs \cite{roth2023frustrated}.

\paragraph{SCALE and sparse updates.}
SCALE (Sparse Convolutional Ansatz for Lattice Electrons) uses one residual convolution without layer normalization \cite{ba2016layer},
\begin{equation}
\mathbf h_i^{(1)}=\mathbf h_i^{(0)}+\sigma\!\left[\mathcal C^{(0)}(\mathbf h^{(0)})_i\right],
\label{eq:scale-block}
\end{equation}
followed by Eqs.~\eqref{eq:site-mlp}-\eqref{eq:backflow}. Changing a few occupations modifies only the features within their convolutional neighborhoods; the remaining orbital rows can be reused from a cache. With consistent occupied-row labels, the updated Slater matrix is $\Phi'=\Phi+UV^{\mathsf T}$, where $U,V\in\mathbb C^{N\times r}$ and $r$ counts the affected rows. $U$ selects these rows and $V^{\mathsf T}$ contains their changes. The matrix determinant lemma and Sherman-Morrison-Woodbury identity give
\begin{align}
B_r&=I_r+V^{\mathsf T}\Phi^{-1}U,\qquad
\frac{\det\Phi'}{\det\Phi}=\det B_r,\nonumber\\
(\Phi')^{-1}&=\Phi^{-1}-\Phi^{-1}UB_r^{-1}V^{\mathsf T}\Phi^{-1}.
\label{eq:woodbury}
\end{align}
Here $I_r$ is the $r\times r$ identity, and $\mathsf T$ denotes the algebraic transpose. Only an $r\times r$ matrix must be factorized; updating the cached inverse costs $\mathcal O(N^2r+r^3)$ rather than a fresh $\mathcal O(N^3)$ factorization. At fixed density, kernel size, and number of changed occupations, $r$ stays bounded as the lattice grows. Together with local feature recomputation, this reduces the determinant-update cost of a sweep of $\mathcal O(N)$ local moves from $\mathcal O(N^4)$ to $\mathcal O(N^3)$ \cite{gu_pareto_2026}.

\paragraph{ACE and deep convolutional features.}
ACE (Accurate Convolutional ansatz for lattice Electrons) replaces Eq.~\eqref{eq:scale-block} by a deep stack,
\begin{equation}
\mathbf h_i^{(\ell+1)}=\operatorname{LN}_{\ell}\!\left(
\mathbf h_i^{(\ell)}+\sigma\!\left[\mathcal C^{(\ell)}(\mathbf h^{(\ell)})_i\right]\right),
\qquad \ell=0,\ldots,L_{\mathrm C}-1.
\label{eq:ace-block}
\end{equation}
Layer normalization \cite{ba2016layer} is applied after the residual addition, independently at each site:
\begin{equation}
\begin{aligned}
{[\operatorname{LN}_{\ell}(\mathbf z_i)]}_\alpha
&=\gamma_{\ell\alpha}\frac{z_{i\alpha}-\bar z_i}{\sqrt{v_i+\epsilon_{\mathrm{LN}}}}+\beta_{\ell\alpha},\\
\bar z_i&=\frac{1}{d_h}\sum_\alpha z_{i\alpha},\qquad
v_i=\frac{1}{d_h}\sum_\alpha(z_{i\alpha}-\bar z_i)^2.
\end{aligned}
\label{eq:layernorm}
\end{equation}
Here $\bar z_i$ and $v_i$ are the mean and variance across channels at site $i$; $\gamma_{\ell\alpha}$ and $\beta_{\ell\alpha}$ are the trainable scale and offset. ACE uses $\epsilon_{\mathrm{LN}}=10^{-4}$. Each layer enlarges the receptive field before the same MLP and orbital output are applied. A sufficiently deep stack allows a local occupation change to affect features throughout the lattice, and the ACE implementation reevaluates the full network rather than using SCALE's sparse updates. SCALE and ACE therefore trade strict update locality for deeper, more expressive backflow features. The Hubbard benchmarks in Ref.~\cite{gu_pareto_2026} report a more favorable energy-runtime trade-off for ACE than for the Transformer-backflow architectures of Ref.~\cite{gu_hubbard_2025} in the systems and configurations compared.

\paragraph{CNN-MPS for spins.}
For spins, the local-state label is $q_i=n_i\in\{0,1\}$, with $n_i=0$ and $1$ denoting spin down and spin up, respectively; $\mathbf n=(n_1,\ldots,n_{N_s})$ specifies the spin configuration. The CNN-MPS implementation initializes its features with trainable vectors $\mathbf e_q,\mathbf p_i\in\mathbb R^{d_h}$: $\mathbf e_q$ embeds local state $q$ and is shared across sites, while $\mathbf p_i$ encodes the position of site $i$,
\begin{equation}
\mathbf h_i^{(0)}=\mathbf e_{q_i}+\mathbf p_i.
\label{eq:cnnmps-embedding}
\end{equation}
It then uses the residual convolutional stack in Eq.~\eqref{eq:ace-block}, with layer normalization at every site, and the site-wise MLP in Eq.~\eqref{eq:site-mlp}. In this architecture, each layer's normalization scale and offset are scalars shared across channels, and $\epsilon_{\mathrm{LN}}=10^{-5}$. Let $d_g$ be the number of output features of the MLP, so $g_{i\mu}(\mathbf n)$ is component $\mu=1,\ldots,d_g$ of $\mathbf g_i(\mathbf n)$. A tensor-valued linear output replaces the orbital map, producing a matrix $A_i^{(r),q}(\mathbf n)\in\mathbb R^{D\times D}$ for each local spin state $q$ and head $r$,
\begin{equation}
[A_i^{(r),q}(\mathbf n)]_{ab}
=\sum_{\mu=1}^{d_g} T^{(r),q}_{ab\mu}g_{i\mu}(\mathbf n),
\qquad a,b=1,\ldots,D.
\label{eq:mps-output}
\end{equation}
Here $a$ and $b$ are the left and right MPS bond indices, and $D$ is the bond dimension. The index $r$ labels one of the $N_{\mathrm{head}}$ MPS heads. The trainable output weights $T^{(r),q}_{ab\mu}$ are shared across sites. Selecting $q=n_i$ and contracting the bond indices with periodic MPS boundary conditions gives, before lattice-symmetry projection,
\begin{equation}
\psi_\theta(\mathbf n)
=\sum_{r=1}^{N_{\mathrm{head}}}\operatorname{Tr}\left[
\prod_{i=1}^{N_s}A_i^{(r),n_i}(\mathbf n)\right].
\label{eq:cnnmps}
\end{equation}
The ordered matrix product follows $i=xL_y+y+1$, with lattice coordinates $x=0,\ldots,L_x-1$ and $y=0,\ldots,L_y-1$; $\operatorname{Tr}$ contracts the last MPS bond with the first. Each head supplies a separate set of output weights while sharing the neural features, and the contracted amplitudes are summed. Because these tensors depend on occupations beyond their own site, this is a backflow MPS, not an ordinary MPS with fixed local tensors. In the neural tensor-network interpretation of Ref.~\cite{fan_disentangling_2026}, the convolutional map acts as a generally nonunitary disentangler and the bond dimension $D$ controls the tensor-network representation of the remaining correlations.

\subsection{Sampling and ground-state optimization}
\label{sec:optimization}

\paragraph{Metropolis-Hastings sampling.}
A Markov chain samples configurations from $p_\theta$ \cite{metropolis1953,hastings1970}. LaQX constructs proposals from the off-diagonal terms of the Hamiltonian: an allowed hopping or spin-exchange term generates a candidate configuration $\mathbf n'$ from $\mathbf n$. A proposal $\mathbf n'\sim q(\mathbf n'|\mathbf n)$ is accepted with probability
\begin{equation}
a(\mathbf n\to\mathbf n')=\min\!\left\{1,
\frac{|\psi_\theta(\mathbf n')|^2}{|\psi_\theta(\mathbf n)|^2}
\frac{q(\mathbf n|\mathbf n')}{q(\mathbf n'|\mathbf n)}\right\}.
\label{eq:metropolis}
\end{equation}
Otherwise the chain remains at $\mathbf n$. Including the forward-reverse proposal ratio accommodates asymmetric proposals while satisfying detailed balance with the Born distribution. Local occupation moves preserve the sampled particle-number or magnetization sector.

\paragraph{Stochastic reconfiguration.}
LaQX provides Adam \cite{kingma2015adam}, SPRING \cite{goldshlager2024spring}, and MARCH \cite{gu_hubbard_2025}. Adam uses the energy-gradient covariance in Eq.~\eqref{eq:energy-gradient} to construct adaptive first- and second-moment updates. The latter two methods are derived from stochastic reconfiguration (SR) \cite{sorella2001sr}: instead of descending in the Euclidean geometry of the parameters, SR projects an imaginary-time step onto the tangent space of the variational state.

Let $\theta=(\theta_1,\ldots,\theta_P)$ contain the $P$ real trainable coordinates, with indices $a,b=1,\ldots,P$. Brackets denote averages over the Born distribution $p_\theta$, and $*$ denotes complex conjugation.

Using the logarithmic derivatives $O_a$ defined above, introduce the centered quantities $\Delta O_a=O_a-\langle O_a\rangle$ and $\Delta E_{\mathrm{loc}}=E_{\mathrm{loc}}-\langle E_{\mathrm{loc}}\rangle$. The derivative covariance matrix $S$ and energy-covariance vector $F$ have entries
\begin{equation}
S_{ab}=\langle\Delta O_a^*\Delta O_b\rangle,\qquad
F_a=\langle\Delta O_a^*\Delta E_{\mathrm{loc}}\rangle.
\label{eq:covariances}
\end{equation}
At optimization step $k$, let $d=(d_1,\ldots,d_P)\in\mathbb R^P$ be a trial parameter-update direction, with $d_a$ specifying the change in coordinate $\theta_a$ before step-size scaling. All averages are evaluated at the current parameters $\theta_k$. Fitting the normalized imaginary-time derivative $-(H-E)|\psi_\theta\rangle$ gives the minimizing direction $d_{\mathrm{SR}}$ and parameter update
\begin{align}
d_{\mathrm{SR}}&=\operatorname*{arg\,min}_{d\in\mathbb R^P}
\left\{\left\langle\left|\sum_{a=1}^{P}\Delta O_a d_a+\Delta E_{\mathrm{loc}}\right|^2\right\rangle+\lambda\|d\|_2^2\right\},\nonumber\\
(\mathrm{Re}\,S+\lambda I_P)d_{\mathrm{SR}}&=-\mathrm{Re}\,F,\qquad
\theta_{k+1}=\theta_k+\eta_k d_{\mathrm{SR}}.
\label{eq:sr}
\end{align}
The real parts arise from minimizing over real parameter changes $d$, even when the wavefunction and its logarithmic derivatives are complex. Here $\lambda>0$ is the damping parameter, $I_P$ is the $P\times P$ identity matrix, and $\|\cdot\|_2$ is the Euclidean norm. The step size is $\eta_k$.

For a batch of $B$ configurations $\mathbf n_\nu$, with sample index $\nu=1,\ldots,B$, let $\overline O_a$ and $\overline E$ denote the sample means of $O_a$ and $E_{\mathrm{loc}}$, respectively. Define the centered derivative matrix $J$ and target vector $\varepsilon$ by
\begin{equation}
J_{\nu a}=\frac{O_a(\mathbf n_\nu)-\overline O_a}{\sqrt B},\qquad
\varepsilon_\nu=-\frac{E_{\mathrm{loc}}(\mathbf n_\nu)-\overline E}{\sqrt B}.
\label{eq:sample-derivatives}
\end{equation}
For a real wavefunction set $X=J$ and $e=\varepsilon$. For a complex wavefunction with real trainable coordinates, stack the two components, $X=(\mathrm{Re}\,J;\mathrm{Im}\,J)$ and $e=(\mathrm{Re}\,\varepsilon;\mathrm{Im}\,\varepsilon)$. Then $X^{\mathsf T}X$ estimates $\mathrm{Re}\,S$ and $X^{\mathsf T}e$ estimates $-\mathrm{Re}\,F$. The damped SR direction has the equivalent forms
\begin{equation}
d_{\mathrm{SR}}=(X^{\mathsf T}X+\lambda I_P)^{-1}X^{\mathsf T}e
=X^{\mathsf T}(XX^{\mathsf T}+\lambda I)^{-1}e.
\label{eq:sr-dual}
\end{equation}
Here $I$ is the identity in sample space, whose dimension is $B$ for real states and $2B$ for complex states. The matrix $X$ has $P$ columns and one row per sample-space component, and $e$ has the corresponding length. The second form reduces the solve to sample space, as in minimum-step stochastic reconfiguration (MinSR) and related formulations \cite{chen2024minsr,rende2024identity}.

In the implementation, local energies are clipped before centering at the sample median plus or minus a chosen multiple of the mean absolute deviation from that median. Real and imaginary parts are clipped separately. Reported energies and energy variances per site use the unclipped local energies.

\paragraph{SPRING.}
Subsampled Projected-Increment Natural Gradient Descent (SPRING) retains the preceding unscaled direction rather than selecting a minimum-norm solution independently for each batch \cite{goldshlager2024spring}. Subscripts $k$ on $X_k$ and $e_k$ denote the sample arrays at optimization step $k$. Let $d_{k-1}$ be the stored direction, $\mu$ the momentum decay coefficient, and $m_k=\mu d_{k-1}$ the decayed direction. SPRING obtains the new unscaled direction $d_k$ by solving
\begin{equation}
d_k=\operatorname*{arg\,min}_{d\in\mathbb R^P}\left\{\|X_kd-e_k\|_2^2+\lambda\|d-m_k\|_2^2\right\}.
\label{eq:spring-objective}
\end{equation}
Its sample-space expression is
\begin{equation}
d_k=m_k+X_k^{\mathsf T}
\left(X_kX_k^{\mathsf T}+\lambda I+R_k\right)^{-1}
(e_k-X_km_k).
\label{eq:spring}
\end{equation}
Thus the new batch corrects the part of the previous direction inconsistent with the current imaginary-time target; this differs from simply adding momentum to an independent SR update. The extra stabilizer is $R_k=\|T_k\|_{\mathrm F}\Pi_0$, where $T_k=X_kX_k^{\mathsf T}$ is the sample-space Gram matrix and $\|\cdot\|_{\mathrm F}$ is the Frobenius norm. For real states, $\Pi_0=\mathbf 1_B\mathbf 1_B^{\mathsf T}/B$ projects onto the constant sample vector, with $\mathbf 1_B$ the length-$B$ vector of ones. For complex states it projects separately onto the constant vectors of the real and imaginary sample blocks. Centering gives $X_k^{\mathsf T}\Pi_0=0$ and $\Pi_0e_k=0$, so this term does not change the exact-arithmetic solution of Eq.~\eqref{eq:spring-objective}; it stabilizes the numerical solve in these null directions.

\paragraph{MARCH.}
The Moment-Adaptive ReConfiguration Heuristic (MARCH) extends this construction by weighting parameter directions according to a second-moment accumulator \cite{gu_hubbard_2025}. In the notation of Eq.~\eqref{eq:sample-derivatives}, the minimization objective is
\begin{equation}
d_k=\operatorname*{arg\,min}_{d\in\mathbb R^P}
\left\{\frac{1}{\lambda}\|X_kd-e_k\|_2^2
+\left\|\operatorname{diag}(v_{k-1})^{1/4}(d-m_k)\right\|_2^2\right\}.
\label{eq:march-objective}
\end{equation}
Here $m_k=\mu d_{k-1}$ is the decayed previous direction and $v_{k-1}$ is the second-moment accumulator, with one positive component per parameter. The operation $\operatorname{diag}(v)$ places the components of $v$ on a diagonal matrix, and its fourth root acts on those entries. Large accumulator components increase the penalty for changing the corresponding parameters. The implementation optionally rescales the accumulator to $\widetilde v_{k-1}$ and adds a small positive $\epsilon$ to form the diagonal penalty metric $D_k$:
\begin{equation}
\begin{aligned}
\widetilde v_{k-1}&=
\begin{cases}
\sqrt P\,v_{k-1}/\|v_{k-1}\|_2,&\text{with accumulator normalization},\\
v_{k-1},&\text{without accumulator normalization},
\end{cases}\\
D_k&=\operatorname{diag}\!\left(\sqrt{\widetilde v_{k-1}}+\epsilon\right).
\end{aligned}
\label{eq:march-metric}
\end{equation}
Square roots act component-wise, and accumulator normalization is enabled by default. With the penalty in Eq.~\eqref{eq:march-objective} evaluated as $(d-m_k)^{\mathsf T}D_k(d-m_k)$, the update is
\begin{align}
\delta_k&=D_k^{-1}X_k^{\mathsf T}
\left(X_kD_k^{-1}X_k^{\mathsf T}+\lambda I+R_k\right)^{-1}(e_k-X_km_k),\nonumber\\
d_k&=m_k+\delta_k,\qquad
v_k=\beta_2\widetilde v_{k-1}+\delta_k\odot\delta_k.
\label{eq:march}
\end{align}
Here $\delta_k$ is the correction to the decayed direction, $\beta_2$ is the accumulator decay coefficient, and $\odot$ denotes component-wise multiplication. The stabilizer is $R_k=\|X_kD_k^{-1}X_k^{\mathsf T}\|_{\mathrm F}\Pi_0$. The initial direction is zero and the initial accumulator has equal positive entries.

Both implementations store $d_k$ before step-length normalization. With $\ell_k$ denoting the scheduled Euclidean length of the parameter change, they apply
\begin{equation}
\theta_{k+1}=\theta_k+\ell_k\frac{d_k}{\|d_k\|_2}.
\label{eq:normalized-update}
\end{equation}
Let $\ell_{\mathrm{base}}$ be the base step length. The schedules are
\begin{equation}
\begin{aligned}
\ell_k^{\mathrm{SPRING}}&=\frac{\ell_{\mathrm{base}}}{1+k/k_0},\\
\ell_k^{\mathrm{MARCH}}&=\ell_{\mathrm{base}}\min\!\left\{\frac{k}{k_{\mathrm w}},\frac{1}{1+\max(k-k_{\mathrm s},0)/k_0}\right\}.
\end{aligned}
\label{eq:optimizer-schedules}
\end{equation}
Here $k_0$ is the decay timescale, $k_{\mathrm w}$ the warmup duration, and $k_{\mathrm s}$ the decay-start step.

\subsection{Natural Excited States}
\label{sec:nes-algorithm}

The natural excited states (NES) method turns the calculation of the lowest $K$ states into a ground-state variational problem on the tensor product of $K$ replicas of the physical Hilbert space \cite{nes_science}. Let $\mathcal N=(\mathbf n_1,\ldots,\mathbf n_K)$ contain one many-body configuration per replica. Given $K$ trainable state functions $\psi_a$, define the matrix with replica index $b$ labeling rows and state index $a$ labeling columns, where $a,b=1,\ldots,K$,
\begin{equation}
M_{ba}(\mathcal N)=\psi_a(\mathbf n_b),\qquad
\Psi(\mathcal N)=\det M(\mathcal N).
\label{eq:nes}
\end{equation}
The expanded Hamiltonian is the Kronecker sum of $K$ copies of the physical Hamiltonian $H$,
\begin{equation}
\begin{aligned}
\mathcal H&=\bigoplus_{b=1}^K H=\sum_{b=1}^K H^{(b)},\\
H^{(b)}&=I^{\otimes(b-1)}\otimes H\otimes I^{\otimes(K-b)}.
\end{aligned}
\label{eq:nes-hamiltonian}
\end{equation}
Here $\bigoplus$ denotes the Kronecker sum, $\otimes$ the tensor product, and $I$ the identity on one replica, with $I^{\otimes0}=1$. Thus $H^{(b)}$ acts only on replica $b$. The determinant makes $\Psi$ antisymmetric under exchange of replicas. If $E_1\leq E_2\leq\cdots$ are the eigenvalues of $H$, the lowest eigenvalue of $\mathcal H$ in this antisymmetric sector is the sum of the lowest $K$ physical energies. The variational objective is therefore
\begin{equation}
\mathcal E=\frac{\langle\Psi|\mathcal H|\Psi\rangle}{\langle\Psi|\Psi\rangle}
\geq\sum_{a=1}^K E_a.
\label{eq:nes-objective}
\end{equation}
Linearly dependent state functions make the determinant identically zero, excluding collapse onto fewer than $K$ independent states without pairwise orthogonality penalties.

LaQX samples $p_{\mathrm{NES}}(\mathcal N)=|\det M|^2/\sum_{\mathcal N'}|\det M(\mathcal N')|^2$. A Metropolis-Hastings proposal changes one replica and uses the ratio of these determinants in Eq.~\eqref{eq:metropolis}. Define the matrix $\mathcal B_{ba}(\mathcal N)=(H\psi_a)(\mathbf n_b)$. Acting with $H^{(b)}$ replaces row $b$ of $M$ by the corresponding row of $\mathcal B$; summing these row replacements gives
\begin{equation}
\begin{aligned}
\mathsf E_{\mathrm{loc}}(\mathcal N)&=M^{-1}\mathcal B,\qquad
\mathcal E=\left\langle\operatorname{Tr}\mathsf E_{\mathrm{loc}}\right\rangle_{p_{\mathrm{NES}}},\\
\frac{(\mathcal H\Psi)(\mathcal N)}{\Psi(\mathcal N)}&=\operatorname{Tr}\mathsf E_{\mathrm{loc}}(\mathcal N).
\end{aligned}
\label{eq:nes-local-matrix}
\end{equation}
Consequently, the same energy-gradient and optimizer formulas apply to $\log\Psi$ and the scalar trace. ACE-NES supplies the individual functions using a shared ACE feature extractor and $K$ state-resolved orbital outputs. Each $\psi_a$ is itself a fermionic determinant over occupied orbitals; the additional determinant in Eq.~\eqref{eq:nes} acts on whole many-body states, not on individual particles.

After optimization, the full energy matrix is accumulated rather than only its trace. Let $G_{ab}=\langle\psi_a|\psi_b\rangle$ and $\mathsf H_{ab}=\langle\psi_a|H|\psi_b\rangle$. For linearly independent functions, its exact expectation satisfies \cite{nes_science}
\begin{equation}
\begin{aligned}
\overline{\mathsf E}&=\langle\mathsf E_{\mathrm{loc}}\rangle_{p_{\mathrm{NES}}}=G^{-1}\mathsf H,\\
\overline{\mathsf E}\,c_a&=\mathcal E_a c_a
\ \Longleftrightarrow\ \mathsf Hc_a=\mathcal E_a Gc_a.
\end{aligned}
\label{eq:nes-diagonalization}
\end{equation}
Diagonalizing the mean matrix gives the Ritz energies $\mathcal E_a$ and coefficient vectors $c_a$ for the corresponding states $\sum_b(c_a)_b\psi_b$.

\subsection{Green's Function Monte Carlo projection}
\label{sec:gfmc-algorithm}

Lattice fixed-node Green's Function Monte Carlo (GFMC) uses imaginary-time propagation to find the ground state of an effective Hamiltonian $H^{\mathrm{FN}}$ constructed from a guiding state \cite{trivedi1990gfmc,sorella1998gfmc}. The neural parameters remain fixed and a weighted walker population represents the evolving state. On configurations where $\psi_G$ is nonzero, introduce the importance-sampled matrix
\begin{equation}
h_{\mathbf n'\mathbf n}=\frac{\psi_G(\mathbf n')}{\psi_G(\mathbf n)}H_{\mathbf n'\mathbf n},\qquad
H_{\mathbf n'\mathbf n}=\langle\mathbf n'|H|\mathbf n\rangle.
\label{eq:gfmc-importance}
\end{equation}
The imaginary-time distribution is $f(\mathbf n,\tau)=\psi_G(\mathbf n)\psi(\mathbf n,\tau)$. Off-diagonal $h_{\mathbf n'\mathbf n}>0$ would produce negative transition weights in a short-time propagator. The fixed-node construction removes these transitions and transfers their contributions to the diagonal \cite{tenhaaf1995fixednode}:
\begin{equation}
\widetilde H^{\mathrm{FN}}_{\mathbf n'\mathbf n}=
\begin{cases}
h_{\mathbf n'\mathbf n},&\mathbf n'\ne\mathbf n,\ h_{\mathbf n'\mathbf n}<0,\\
0,&\mathbf n'\ne\mathbf n,\ h_{\mathbf n'\mathbf n}\geq0,\\
H_{\mathbf n\mathbf n}+\displaystyle\sum_{\mathbf m\ne\mathbf n}\max(h_{\mathbf m\mathbf n},0),&\mathbf n'=\mathbf n.
\end{cases}
\label{eq:fixed-node}
\end{equation}
The matrix $\widetilde H^{\mathrm{FN}}$ represents $H^{\mathrm{FN}}$ after importance sampling. The fixed-node correction preserves the guiding-state local energy, so the walker weights and mixed energy estimate use
\begin{equation}
E_{\mathrm{loc}}^G(\mathbf n)
=\frac{(H^{\mathrm{FN}}\psi_G)(\mathbf n)}{\psi_G(\mathbf n)}
=\frac{(H\psi_G)(\mathbf n)}{\psi_G(\mathbf n)}.
\label{eq:gfmc-local-energy}
\end{equation}
Together with the positive semidefiniteness of $H^{\mathrm{FN}}-H$, this gives the variational bound \cite{tenhaaf1995fixednode}
\begin{equation}
E_0\leq E_0^{\mathrm{FN}}\leq E[\psi_G],\qquad
E[\psi_G]=\frac{\langle\psi_G|H|\psi_G\rangle}{\langle\psi_G|\psi_G\rangle}.
\label{eq:gfmc-bound}
\end{equation}
Here $E_0$ and $E_0^{\mathrm{FN}}$ are the ground-state energies of $H$ and $H^{\mathrm{FN}}$, respectively. The fixed-node bias vanishes when the guiding function is an exact ground state.

For a sufficiently small uniform step $\Delta\tau$, the nonnegative linear propagator is
\begin{equation}
\widetilde G=I-\Delta\tau(\widetilde H^{\mathrm{FN}}-E_T I),
\label{eq:gfmc-propagator}
\end{equation}
where $E_T$ is a reference energy and $I$ is the identity in configuration space. LaQX limits the step at each walker to avoid a negative diagonal entry. Writing $d_{\mathbf n}=\widetilde H^{\mathrm{FN}}_{\mathbf n\mathbf n}$ and denoting the requested step by $\Delta\tau_0$, the implemented rule is
\begin{equation}
\Delta\tau_{\mathbf n}=
\begin{cases}
(d_{\mathbf n}-E_T)^{-1},&d_{\mathbf n}-E_T>1/\Delta\tau_0,\\
\Delta\tau_0,&\text{otherwise}.
\end{cases}
\label{eq:gfmc-step}
\end{equation}
For each walker, the transition and weight update are then
\begin{align}
\widetilde G_{\mathbf n'\mathbf n}
&=\delta_{\mathbf n'\mathbf n}-\Delta\tau_{\mathbf n}
\left(\widetilde H^{\mathrm{FN}}_{\mathbf n'\mathbf n}-E_T\delta_{\mathbf n'\mathbf n}\right),\nonumber\\
b(\mathbf n)&=\sum_{\mathbf n'}\widetilde G_{\mathbf n'\mathbf n}
=1+\Delta\tau_{\mathbf n}\left[E_T-E_{\mathrm{loc}}^G(\mathbf n)\right],\nonumber\\
p(\mathbf n'|\mathbf n)&=\widetilde G_{\mathbf n'\mathbf n}/b(\mathbf n),\qquad
w'=w\,b(\mathbf n).
\label{eq:gfmc-transition}
\end{align}
Here $\delta$ is the Kronecker delta, $b$ is the column sum, and $w$ and $w'$ are the walker weights before and after propagation. This update samples a propagator column directly, including its diagonal stay-put contribution. For $N_w$ walkers with configurations $\mathbf n_p$ and weights $w_p$, the mixed energy estimate is
\begin{equation}
E_{\mathrm{mix}}=\frac{\sum_{p=1}^{N_w}w_pE_{\mathrm{loc}}^G(\mathbf n_p)}{\sum_{p=1}^{N_w}w_p}.
\label{eq:gfmc-estimator}
\end{equation}

\subsection{Adiabatic evolution}
\label{sec:adiabatic-algorithm}

Adiabatic evolution is approximated by a sequence of variational optimizations for nearby Hamiltonians $H(\varphi_j)$, where $j$ labels successive values of a control parameter $\varphi$. Starting from a prepared state, the optimized parameters at $\varphi_j$ initialize the calculation at $\varphi_{j+1}$:
\begin{equation}
\theta_{j+1}=\operatorname*{arg\,min}_{\theta\leftarrow\theta_j}
\frac{\langle\psi_\theta|H(\varphi_{j+1})|\psi_\theta\rangle}{\langle\psi_\theta|\psi_\theta\rangle}.
\label{eq:adiabatic-continuation}
\end{equation}
The arrow denotes initialization at $\theta_j$. Small parameter increments and a fixed symmetry sector allow the optimization to follow a continuous low-energy branch.

In the charge-pumping application \cite{hofstadter_pumping_2026}, $\varphi$ is a boundary twist increased from zero to $2\pi$. In the Hofstadter implementation, the $x$-direction hopping takes the uniform-gauge form
\begin{equation}
H_x(\varphi)=-\sum_{x,y}\left(
e^{i\varphi/L_x}c_{x+1,y}^{\dagger}c_{x,y}
+\mathrm{H.c.}\right).
\label{eq:hofstadter-twist}
\end{equation}
Thus hopping in the positive $x$ direction carries $e^{i\varphi/L_x}$, and reverse hopping carries its complex conjugate. The accumulated phase around the periodic $x$ direction is $e^{i\varphi}$, so increasing $\varphi$ by $2\pi$ inserts one flux quantum through the torus. Each step is followed by a polarization measurement. Continuing the low-energy momentum sectors separately gives the charge-pumping response described in Sec.~\ref{sec:charge-pumping}.

\subsection{Time-dependent VMC and parameter-efficient evolution}
\label{sec:tvmc-algorithm}

\paragraph{ACE-PEFT.}
Parameter-efficient fine-tuning (PEFT) restricts time evolution to the final affine orbital map of a prepared ACE state, adapting the neural-basis tVMC principle of Ref.~\cite{fu_nbtvmc_2026}. Let $d_g$ be the number of features in $\mathbf g_i$ from Eq.~\eqref{eq:site-mlp}. Define $\widetilde g_{i0}=1$ to include the output bias and $\widetilde g_{i\mu}=g_{i\mu}$ for $\mu=1,\ldots,d_g$, so that
\begin{equation}
\begin{aligned}
M_{(i,s),j}(\mathbf n,t)&=\sum_{\mu=0}^{d_g}W_{sj\mu}(t)\widetilde g_{i\mu}(\mathbf n;\eta_0),\\
\psi_{W(t)}(\mathbf n)&=\det\!\left[M_{I_a(\mathbf n),j}(\mathbf n,t)\right]_{a,j=1}^{N}.
\end{aligned}
\label{eq:ace-peft}
\end{equation}
Here $i$, $s$, and $j$ label the site, spin channel, and orbital, and $I_a(\mathbf n)$ selects the occupied spin orbitals as in Eq.~\eqref{eq:backflow}. The embedding, convolutional, layer-normalization, and MLP parameters $\eta_0$ remain fixed. Only the complex coefficients $W(t)$, including the bias coefficients $W_{sj0}$, evolve. Freezing the nonlinear feature extractor reduces the number of dynamical parameters while retaining its configuration-dependent correlations.

\paragraph{Projected real-time dynamics.}
Time-dependent VMC approximates $i\partial_t|\psi\rangle=H|\psi\rangle$, with $\hbar=1$, by evolution within a variational manifold \cite{carleo2012localization,schmitt2020dynamics}. Collect the $P$ evolved coefficients of $W$ into $\theta\in\mathbb C^P$. The ACE-PEFT amplitude is holomorphic in these coefficients, and its logarithmic derivatives $O_a=\partial_{\theta_a}\log\psi_\theta$, $a=1,\ldots,P$, are evaluated by automatic differentiation. With the centered quantities and Born-distribution averages of Sec.~\ref{sec:optimization}, the time-dependent variational principle (TDVP) determines the parameter velocity $\dot\theta=d\theta/dt$ by
\begin{equation}
\dot\theta
=\operatorname*{arg\,min}_{v\in\mathbb C^P}
\left\langle\left|\sum_{a=1}^{P}\Delta O_a v_a+i\Delta E_{\mathrm{loc}}\right|^2\right\rangle.
\label{eq:tdvp-objective}
\end{equation}
Here $v$ is a trial complex parameter velocity. Minimization gives
\begin{equation}
\sum_{b=1}^{P} S_{ab}\dot\theta_b=-iF_a,
\label{eq:tdvp}
\end{equation}
where $S$ and $F$ have the covariance form of Eq.~\eqref{eq:covariances}, evaluated with the complex-coordinate derivatives.

\paragraph{Regularization and integration.}
To regularize small eigenvalues of the sampled covariance matrix, LaQX either solves $(S+\lambda I_P)\dot\theta=-iF$ by Cholesky factorization, with damping $\lambda>0$ and $P\times P$ identity $I_P$, or diagonalizes $S=V\operatorname{diag}(s_\nu)V^\dagger$ and uses
\begin{equation}
\begin{aligned}
\dot\theta&=-iV\operatorname{diag}[f(s_\nu)]V^\dagger F,\\
f(s_\nu)&=
\begin{cases}
\displaystyle\frac{1}{s_\nu}\frac{1}{1+(c/|s_\nu/s_{\max}|)^6},&|s_\nu/s_{\max}|>\epsilon_{\mathrm{eig}},\\
0,&\text{otherwise}.
\end{cases}
\end{aligned}
\label{eq:tdvp-filter}
\end{equation}
Here $s_\nu$ are the eigenvalues, $\nu=1,\ldots,P$, the columns of $V$ are the corresponding orthonormal eigenvectors, and $\dagger$ denotes the Hermitian conjugate. The largest eigenvalue is $s_{\max}$, $c$ is the smooth relative cutoff, and $\epsilon_{\mathrm{eig}}=10^{-14}$ is the numerical threshold.

Writing the regularized velocity as $\mathcal V(\theta,t)$ and $\theta_t=\theta(t)$, the fourth-order Runge-Kutta (RK4) update \cite{butcher1996rungekutta} over a time step $\Delta t$ uses four stage velocities $k_1,\ldots,k_4$:
\begin{align}
k_1&=\mathcal V(\theta_t,t),\nonumber\\
k_2&=\mathcal V(\theta_t+\tfrac{\Delta t}{2}k_1,t+\tfrac{\Delta t}{2}),\nonumber\\
k_3&=\mathcal V(\theta_t+\tfrac{\Delta t}{2}k_2,t+\tfrac{\Delta t}{2}),\nonumber\\
k_4&=\mathcal V(\theta_t+\Delta t k_3,t+\Delta t),\nonumber\\
\theta_{t+\Delta t}&=\theta_t+\frac{\Delta t}{6}(k_1+2k_2+2k_3+k_4).
\label{eq:rk4}
\end{align}
At each stage, walkers sample the intermediate wavefunction and $S$ and $F$ are reestimated. Observables are averaged over the stages with weights $(1,2,2,1)/6$.

\section{Function-oriented software design}
\label{sec:software}

\subsection{Composition of the numerical workflow}

In LaQX, a complete calculation is divided into four scientific components: the variational state, the Hamiltonian and observables, the sampling procedure, and the parameter-update rule. Each component is exposed through a small functional interface, so a complete workflow is assembled by composition rather than through a hierarchy of stateful solver objects. Numerical state is passed explicitly between components. This design permits compilation, vectorization, automatic differentiation, and device parallelism to be applied to the assembled numerical update \cite{jax2018}.

The state layer implements the architectures described in Sec.~\ref{sec:ansatz}, together with the ACE-NES and ACE-PEFT specializations of Secs.~\ref{sec:nes-algorithm} and \ref{sec:tvmc-algorithm}. Independently, the operator layer represents Hamiltonians and observables through the configurations and matrix elements connected to each sampled occupation vector. Sampling and optimization then act on these common representations without depending on the internal construction of a particular ansatz or model.

This separation also determines the parallelization strategy. Batches of sampled configurations are distributed across accelerator devices, while the variational parameters and update state are shared. Ground-state optimization methods reuse the same outer VMC workflow, and measurement, projection, excited-state, and dynamical calculations reuse the same state and operator interfaces. The architecture therefore separates physical definitions from sampling, optimization, and distributed execution while preserving a common numerical core.

\subsection{Locality of changes and LLM-assisted development}
\label{sec:vibe-coding}

Small functional interfaces also support LLM-assisted programming. Reconstructing a large inheritance hierarchy or tracing implicit mutable state across many files consumes additional context and tokens, increasing the cost of a change. In LaQX, many extensions require only the relevant scientific component and its interface to be included in the context.

Table~\ref{tab:extension-locality} summarizes this locality at the level of scientific concepts. A new Hamiltonian defines its connected configurations and matrix elements, after which the existing local-energy and sampling machinery can be reused. A new ansatz supplies initialization and amplitude evaluation without changing the Hamiltonian or optimization methods. Likewise, a new update rule can operate on the common stochastic estimators without depending on the construction of the state.

\begin{table}[t]
\centering
\begin{tabular}{>{\raggedright\arraybackslash}p{0.19\linewidth}>{\raggedright\arraybackslash}p{0.31\linewidth}>{\raggedright\arraybackslash}p{0.40\linewidth}}
\toprule
Extension & Required scientific definition & Reused components \\
\midrule
Variational state & Initialization and amplitudes & Hamiltonians, sampling, and optimization \\
\midrule
Hamiltonian or observable & Connected configurations and matrix elements & Variational states and update methods \\
\midrule
Optimization method & Update from stochastic estimators & State and operator definitions \\
\midrule
Physical workflow & Composition of established components & States, operators, and numerical kernels \\
\bottomrule
\end{tabular}
\caption{Locality of the principal scientific extension paths in LaQX.}
\label{tab:extension-locality}
\end{table}

The example of the altermagnetic Hubbard model in Sec.~\ref{sec:altermagnet} provides a concrete measure of this locality.

\section{Physical applications}
\label{sec:examples}

We now specify the example physical problems used to exercise the code. They cover fermionic and spin ground states, one-body observables, many-body topological responses, low-energy manifolds, and real-time dynamics. Reproducible input specifications and documentation accompany each example in the library.

The square-lattice Hubbard Hamiltonian \cite{hubbard1963} is defined as
\begin{equation}
H_{\mathrm{Hub}}
=
-t \sum_{\langle i,j\rangle,\sigma}
\left(c_{i\sigma}^{\dagger}c_{j\sigma}+c_{j\sigma}^{\dagger}c_{i\sigma}\right)
+
U\sum_i n_{i\uparrow}n_{i\downarrow}.
\label{eq:hubbard}
\end{equation}
Here $c_{i\sigma}^{\dagger}$ creates a fermion of spin $\sigma$ at site $i$, $n_{i\sigma}=c_{i\sigma}^{\dagger}c_{i\sigma}$, $n_i=n_{i\uparrow}+n_{i\downarrow}$, and $\langle i,j\rangle$ denotes nearest-neighbor bonds. We set $t=1$ and consider only nearest-neighbor hopping in this benchmark. Extensions with next-nearest-neighbor hopping $t'$ are used in studies of competing stripe and superconducting orders \cite{roth2025superconductivity,viteritti2026bias,rende2026superconductivity}.

The frustrated spin benchmark uses the square-lattice spin-$1/2$ $J_1$-$J_2$ Heisenberg Hamiltonian
\begin{equation}
H_{J_1J_2}
=
J_1\sum_{\langle i,j\rangle}\mathbf{S}_i\cdot\mathbf{S}_j
+
J_2\sum_{\langle\langle i,j\rangle\rangle}\mathbf{S}_i\cdot\mathbf{S}_j .
\label{eq:j1j2}
\end{equation}
Here $\mathbf S_i$ is the spin operator at site $i$, and $\langle\langle i,j\rangle\rangle$ denotes next-nearest-neighbor bonds. The calculation uses a fixed-magnetization sector.

The interacting Hofstadter examples \cite{hofstadter1976,hofstadter_pumping_2026} use
\begin{equation}
H_{\mathrm{Hof}}
=
-\sum_{x,y}
\left(
c_{x+1,y}^{\dagger}c_{x,y}
+
e^{2\pi i\alpha x}c_{x,y+1}^{\dagger}c_{x,y}
+
\mathrm{h.c.}
\right)
+
V\sum_{\langle i,j\rangle}n_i n_j ,
\label{eq:hofstadter}
\end{equation}
Here $c_{x,y}$ annihilates a spin-polarized fermion at site $(x,y)$, $n_i=c_i^\dagger c_i$, and $\mathrm{h.c.}$ denotes the Hermitian conjugate. The magnetic flux per plaquette is $\alpha$ in units of the flux quantum, and $V$ is the nearest-neighbor density interaction; the hopping amplitude is set to unity. A boundary twist $\varphi$ through the periodic $x$ direction replaces the $x$-direction hopping by Eq.~\eqref{eq:hofstadter-twist}.

The altermagnetic Hubbard model adds sublattice- and direction-dependent diagonal hopping to Eq.~\eqref{eq:hubbard} \cite{alter_prl}:
\begin{equation}
H_{\mathrm{alt}}=H_{\mathrm{Hub}}
-\sum_{\langle\langle i,j\rangle\rangle,\sigma}
t^{(2)}_{ij}\left(c_{i\sigma}^{\dagger}c_{j\sigma}+\mathrm{H.c.}\right).
\label{eq:altermagnetic-hubbard}
\end{equation}
Let $\mathbf r_i=(x_i,y_i)$ be the integer coordinates of site $i$, with A and B denoting sites with odd and even $x_i+y_i$, respectively. The real hopping amplitude $t^{(2)}_{ij}$ depends on the sublattice of site $i$ and the bond displacement $\mathbf r_j-\mathbf r_i$:
\begin{equation}
\begin{array}{c|cc}
 & (1,1) & (1,-1) \\
\hline
\mathrm A & t_- & t_+ \\
\mathrm B & t_+ & t_-
\end{array}
\label{eq:altermagnetic-hoppings}
\end{equation}
Opposite bond orientations have the same amplitude, $t^{(2)}_{ji}=t^{(2)}_{ij}$, and each undirected diagonal bond is counted once in Eq.~\eqref{eq:altermagnetic-hubbard}.

The resulting set of calculations is summarized in Table~\ref{tab:examples}.

\begin{table}[t]
\centering
\begin{tabular}{>{\raggedright\arraybackslash}p{0.18\linewidth}>{\raggedright\arraybackslash}p{0.29\linewidth}>{\raggedright\arraybackslash}p{0.18\linewidth}>{\raggedright\arraybackslash}p{0.235\linewidth}}
\toprule
Protocol & Physics target & Variational state & Main output \\
\midrule
Hubbard & Ground-state VMC and GFMC & ACE, SCALE & Energy traces and projected energy \\
\midrule
Spin & Frustrated $J_1$-$J_2$ model & CNN-MPS & Energy and variance per site \\
\midrule
Altermagnet & Spin-resolved momentum structure & ACE & Density matrix and $\langle n_{\mathbf{k}\sigma}\rangle$ \\
\midrule
NES & Low-lying Hofstadter states & ACE-NES & Energy matrix and spectrum \\
\midrule
Charge pumping & Hofstadter topological response & ACE & Momentum sectors and polarization \\
\midrule
tVMC & Real-time density dynamics & ACE-PEFT & Density time series \\
\bottomrule
\end{tabular}
\caption{Physical systems, variational states, and measured quantities in the documented LaQX calculations.}
\label{tab:examples}
\end{table}

\subsection{Doped Hubbard ground states and GFMC}

Large fermionic ansatzes are initialized through a sequence of variational states and Hamiltonians \cite{gu_hubbard_2025,warm_start_2025}. The reference state is a determinant constructed directly from trainable local orbital embeddings,
\begin{equation}
\begin{aligned}
M^{\mathrm{ref}}_{(i,s),j}(\mathbf n)&=A_{i,q_i,s,j},\\
\psi_{\mathrm{ref}}(\mathbf n;A)
&=\det\!\left[M^{\mathrm{ref}}_{I_a(\mathbf n),j}(\mathbf n)\right]_{a,j=1}^{N}.
\end{aligned}
\label{eq:reference-determinant}
\end{equation}
Here $q_i=n_{i\uparrow}+2n_{i\downarrow}$ labels the local occupation, $s$ is the spin channel, $j$ is the orbital index, and $I_a(\mathbf n)$ labels the occupied spin orbitals as in Eq.~\eqref{eq:backflow}. The entries $A_{i,q_i,s,j}$ are optimized directly, with no convolutional or MLP layers.

Both Hubbard calculations use $1/8$ hole doping. For the periodic $16\times4$ lattice with $N=56$ electrons, this reference state is first optimized at $U=4$ with Adam. It then initializes an ACE state, optimized for $30000$ MARCH steps at $U=8$. For the periodic $16\times8$ lattice with $N=112$, the reference stage includes a stripe-pinning field of strength $0.05$ and charge period eight, targeting a filled stripe \cite{gu_pareto_2026}. The field is removed for the final $20000$-step SCALE optimization at $U=8$. Both optimizations use $4096$ walkers. The ACE state has 12 convolutional layers of width 128; SCALE uses width 256 and four site-wise nonlinear layers of width 512, following the architecture families of Ref.~\cite{gu_pareto_2026}.

\begin{figure}[t]
\centering
\includegraphics[width=0.48\linewidth]{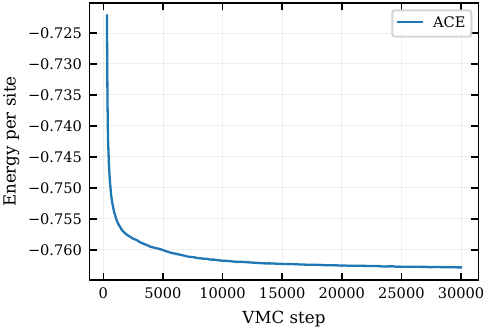}
\includegraphics[width=0.48\linewidth]{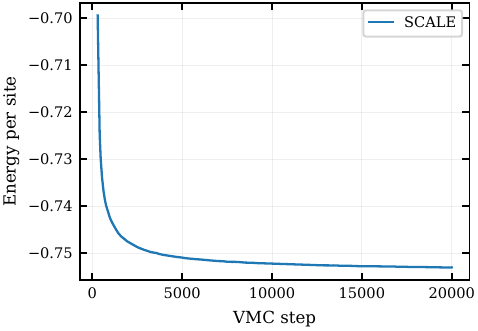}
\caption{VMC energy traces for the doped square-lattice Hubbard model. Left: ACE on a periodic $16\times4$ lattice with $N=56$ electrons. Right: SCALE on a periodic $16\times8$ lattice with $N=112$ electrons. Both final optimizations use $U=8$ and the staged initialization described in the text. Total energies are divided by 64 and 128, respectively, and the curves show trailing 300-step averages.}
\label{fig:hubbard}
\end{figure}

\begin{figure}[t]
\centering
\includegraphics[width=0.96\linewidth]{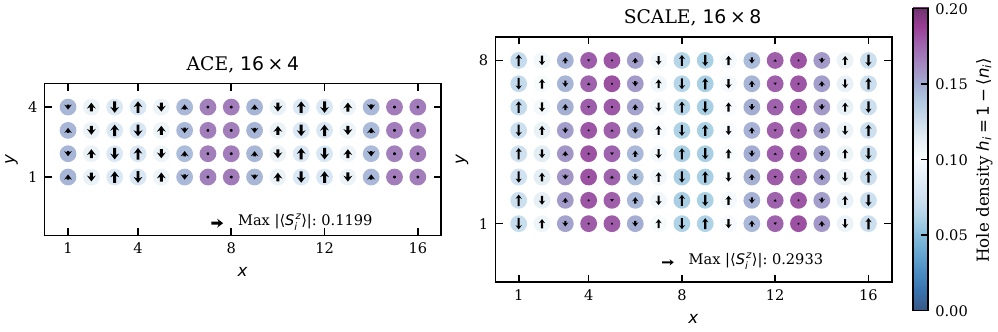}
\caption{Local stripe profiles in the final Hubbard variational states at $U=8$: ACE on $16\times4$ (left) and SCALE on $16\times8$ (right). Circle colors encode the hole density $h_i=1-\langle n_i\rangle$. Vertical arrows encode the unstaggered spin polarization $\langle S_i^z\rangle$, with direction indicating its sign and length proportional to its magnitude within each panel; very small moments are rendered as dots. Both panels use the same color scale, but arrow lengths are scaled separately so that the maximum absolute spin polarization in each panel has the same displayed arrow length. Each reference arrow is labeled with the corresponding maximum absolute spin polarization. No spatial smoothing or symmetry averaging is applied.}
\label{fig:hubbard-stripes}
\end{figure}

The local observables in Fig.~\ref{fig:hubbard-stripes} show the hole density $h_i=1-\langle n_{i\uparrow}+n_{i\downarrow}\rangle$ through site colors and the spin polarization $\langle S_i^z\rangle=\langle n_{i\uparrow}-n_{i\downarrow}\rangle/2$ through vertical arrows. Both states exhibit filled stripes \cite{zhengStripeOrderUnderdoped2017} with a charge-density-wave wavelength of eight sites along the long direction and an antiphase modulation of the staggered magnetization $m_i=(-1)^{x_i+y_i}\langle S_i^z\rangle$.

Evaluation of the final states gives energies per site of $-0.762822(5)$ for ACE on $16\times4$ and $-0.753059(7)$ for SCALE on $16\times8$. Both states are subsequently used as importance-sampling and fixed-node guiding functions for the GFMC algorithm of Sec.~\ref{sec:gfmc-algorithm}, with the Hamiltonians and boundary conditions held fixed. Discarding the first $1000$ of $5000$ projection steps gives energies per site of $-0.76378(2)$ for ACE-guided GFMC and $-0.75628(3)$ for SCALE-guided GFMC. These estimates lower the respective VMC energies by approximately $9.6\times10^{-4}$ and $3.2\times10^{-3}$ per site.

\subsection{Frustrated $J_1$-$J_2$ spin model}

For the spin benchmark we consider the periodic $6\times6$ $J_1$-$J_2$ Heisenberg model defined in Eq.~\eqref{eq:j1j2} at $J_1=1$ and $J_2=0.5$, in the zero-magnetization sector. The variational state is CNN-MPS with MPS bond dimension $D=20$ \cite{fan_disentangling_2026}: the convolutional feature map first reorganizes the correlations of the sampled spin configuration, after which a periodic backflow MPS contracts the configuration-dependent tensors as in Eq.~\eqref{eq:cnnmps}. The calculation applies the Marshall sign transformation \cite{marshall1955} and projection onto the $D_4$ lattice-symmetry sector. The optimization uses $15000$ MARCH steps without second-moment accumulator normalization, corresponding to $\widetilde v_{k-1}=v_{k-1}$ in Eq.~\eqref{eq:march-metric}. The step-length normalization in Eq.~\eqref{eq:normalized-update} remains active. In addition to the energy, we report the local-energy variance per site,
\begin{equation}
\sigma_E^2=\frac{\mathrm{Var}(E_{\mathrm{loc}})}{N_s}
=\frac{\langle |E_{\mathrm{loc}}|^2\rangle-|\langle E_{\mathrm{loc}}\rangle|^2}{N_s},
\qquad N_s=36.
\end{equation}

\begin{figure}[t]
\centering
\includegraphics[width=0.96\linewidth]{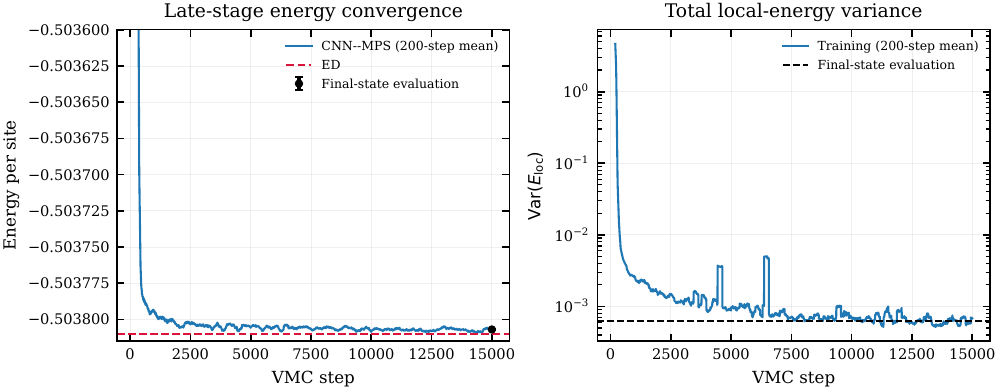}
\caption{CNN-MPS optimization of the periodic $6\times6$ $J_1$-$J_2$ model at $J_2/J_1=0.5$, using MARCH without second-moment accumulator normalization. Left: late-stage convergence of the energy per site $e=\langle H_{J_1J_2}\rangle/36$, with the rounded ED reference $e_{\mathrm{ED}}=-0.503810$ \cite{schulz1996j1j2,fan_disentangling_2026}. Right: local-energy variance per site $\sigma_E^2=\mathrm{Var}(E_{\mathrm{loc}})/36$. Training curves are trailing 200-step averages.}
\label{fig:spin}
\end{figure}

Independent evaluation of the final state gives an energy per site of $e=-0.5038070(2)$ and a local-energy variance per site of $\sigma_E^2=1.7586\times10^{-5}$ in units of $J_1^2$. The measured energy lies approximately $3\times10^{-6}$ per site above the rounded ED reference in Fig.~\ref{fig:spin}.

\subsection{Altermagnetic Hubbard model}
\label{sec:altermagnet}

The altermagnetic calculation tests both a nonstandard kinetic term and an observable beyond the total energy. The diagonal hopping amplitudes in Eq.~\eqref{eq:altermagnetic-hoppings} are $t_+=0.4$ and $t_-=0$. A half-filled periodic $12\times12$ system is first optimized at $U=3$ using the reference determinant ansatz in Eq.~\eqref{eq:reference-determinant}. This occupation-dependent reference state initializes the subsequent ACE optimization at $U=8$. The trained state is then sampled to obtain the fermionic one-body density matrix.

\begin{figure}[t]
\centering
\includegraphics[width=0.58\linewidth]{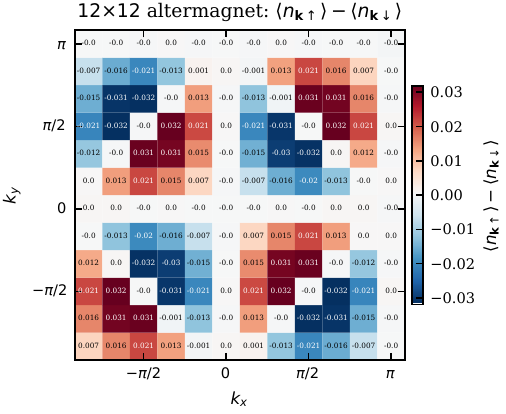}
\caption{Spin-channel difference $\langle n_{\mathbf k\uparrow}\rangle-\langle n_{\mathbf k\downarrow}\rangle$ for the half-filled periodic $12\times12$ altermagnetic Hubbard model at $U=8$. Both momentum distributions are obtained from the sampled one-body density matrix.}
\label{fig:alter}
\end{figure}

Let $c_{\mathbf k\sigma}=N_s^{-1/2}\sum_j e^{-i\mathbf k\cdot\mathbf r_j}c_{j\sigma}$ annihilate a fermion of momentum $\mathbf k$ and spin $\sigma$, where $\mathbf r_j$ is the position of site $j$. The measured momentum distribution is
\begin{equation}
\langle n_{\mathbf{k}\sigma}\rangle
=\langle c_{\mathbf k\sigma}^\dagger c_{\mathbf k\sigma}\rangle=
\frac{1}{N_s}
\sum_{i,j}
e^{i\mathbf{k}\cdot(\mathbf{r}_i-\mathbf{r}_j)}
\langle c_{i\sigma}^{\dagger}c_{j\sigma}\rangle .
\end{equation}
Figure~\ref{fig:alter} shows the difference $\langle n_{\mathbf k\uparrow}\rangle-\langle n_{\mathbf k\downarrow}\rangle$, highlighting the anisotropic spin-dependent momentum structure obtained from this estimator.

\subsection{Hofstadter low-energy spectrum with NES}

\begin{figure}[t]
\centering
\includegraphics[width=0.70\linewidth]{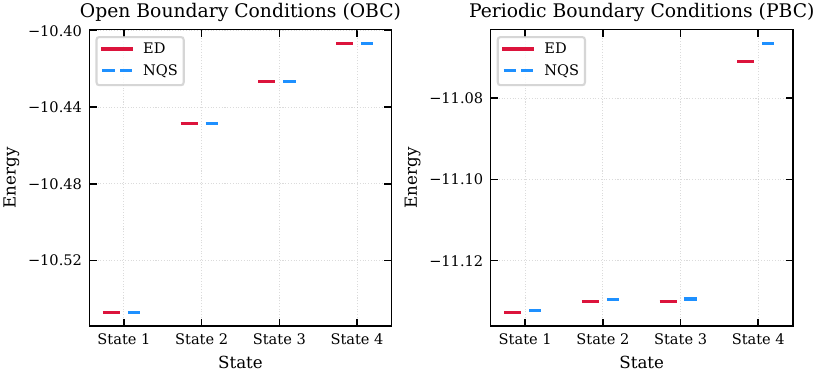}
\caption{Low-energy spectrum obtained with four-state ACE-NES for a $12\times4$ interacting Hofstadter model with four spin-polarized particles, $\alpha=0.25$, and $V=2$, compared with ED. Left: open boundaries. Right: periodic boundaries. The NQS energies are the eigenvalues of the Monte Carlo-averaged energy matrix; eigenvalue error bars are omitted because the stored data do not include the matrix covariances required for uncertainty propagation.}
\label{fig:nes}
\end{figure}

The NES calculation uses a $12\times4$ interacting Hofstadter system with $N=4$ spin-polarized particles, $\alpha=0.25$, and $V=2$. Four state-resolved determinant outputs share the nonlinear ACE feature extractor and are optimized jointly through the auxiliary determinant in Eq.~\eqref{eq:nes}, under either open or periodic boundary conditions. Sampling after optimization gives the $4\times4$ energy matrix; its Monte Carlo mean is diagonalized to resolve the energies represented in the learned subspace. Figure~\ref{fig:nes} compares these eigenvalues with ED.

The open-boundary NES energies are approximately $-10.5473$, $-10.4485$, $-10.4265$, and $-10.4066$, with deviations from ED ranging from $4.6\times10^{-5}$ to $1.0\times10^{-4}$. The finite cluster has a nondegenerate ground state separated from the first excitation by approximately $0.099$.

For periodic boundaries, the energies are approximately $-11.1322$, $-11.1295$, $-11.1295$, and $-11.0665$. The lowest three states form a closely spaced group, while the fourth lies about $0.063$ above the third. The first three deviations from ED are below $6.1\times10^{-4}$, whereas the fourth is about $4.5\times10^{-3}$, indicating a less accurate representation of that state. At filling $\nu=N/(\alpha N_s)=1/3$, this three-state manifold is consistent with the torus ground-state structure of a fractional quantum Hall phase \cite{hofstadter_pumping_2026}. Section~\ref{sec:charge-pumping} examines the associated fractional charge-pumping response.

\subsection{Hofstadter charge pumping}
\label{sec:charge-pumping}

Following the charge-pumping formulation of Ref.~\cite{hofstadter_pumping_2026}, Eq.~\eqref{eq:hofstadter} is defined on a periodic $12\times8$ lattice with $N=8$ spin-polarized particles, $\alpha=0.25$, and $V=2$. In this gauge, the magnetic unit cell spans four sites along $x$. The integers $K_x=0,1,2$ and $K_y=0,\ldots,7$ label momentum sectors for translations by one magnetic unit cell along $x$ and one site along $y$, respectively. For a translation by $(4m,n)$ sites, the symmetry-projection phase is
\begin{equation}
\chi_{K_x,K_y}(m,n)
=\exp\!\left[2\pi i\left(\frac{K_xm}{3}+\frac{K_yn}{8}\right)\right],
\qquad m=0,1,2,\quad n=0,\ldots,7.
\label{eq:hofstadter-sectors}
\end{equation}

\begin{figure}[t]
\centering
\includegraphics[width=0.48\linewidth]{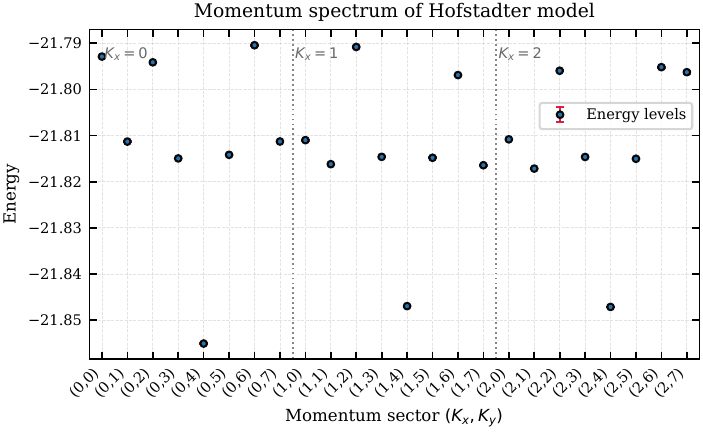}
\includegraphics[width=0.48\linewidth]{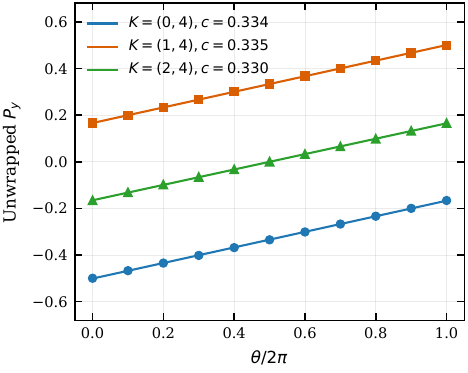}
\caption{Charge pumping in the interacting Hofstadter model on a $12\times8$ torus. Left: spectrum resolved by the integer momentum-sector labels $(K_x,K_y)$ defined in Eq.~\eqref{eq:hofstadter-sectors}, with the three lowest states in $(0,4)$, $(1,4)$, and $(2,4)$. Energy bars show nominal standard errors without an autocorrelation correction. Right: unwrapped dimensionless polarization $P_y$ and separate linear fits versus $\varphi/(2\pi)$ for each continued state; $c$ denotes the fitted slope.}
\label{fig:hofstadter}
\end{figure}

The sector scan identifies three low-energy states at $(K_x,K_y)=(0,4)$, $(1,4)$, and $(2,4)$, as shown in Fig.~\ref{fig:hofstadter} (left). Their measured total energies are $-21.85505(9)$, $-21.84695(9)$, and $-21.84713(9)$, respectively; the next level in the scanned sectors lies at $-21.81715(8)$, about $0.030$ above the top of this manifold. The finite variational splitting within the manifold is approximately $0.0081$. Separate adiabatic-continuation calculations follow these three states using the protocol of Sec.~\ref{sec:adiabatic-algorithm}. The boundary twist $\varphi$ increases from $0$ to $2\pi$, inserting one flux quantum.

At each flux value, the $y$-direction Resta polarization is measured \cite{resta1998},
\begin{equation}
\hat{Z}_y
=
\exp\left(
\frac{2\pi i}{L_y}
\sum_{x,y}y\,\hat{n}_{x,y}
\right),
\qquad
P_y(\varphi)
=
\frac{1}{2\pi}
\mathrm{Im}\log
\langle \hat{Z}_y\rangle_{\varphi}.
\end{equation}
The measured phase is unwrapped before division by $2\pi$. Separate linear fits of $P_y$ against $\varphi/(2\pi)$ give slopes $0.334$, $0.335$, and $0.330$ for the three continued states (Fig.~\ref{fig:hofstadter}, right). Each branch therefore pumps approximately one third of a particle per inserted flux quantum, with a combined response of approximately one particle. This fractional response is consistent with flux insertion connecting states within a topological ground-state manifold \cite{hofstadter_pumping_2026}.

LaQX was also used in a recent study of the triangular-lattice Hofstadter-Hubbard model \cite{gu_spincharge_2026}. There, LaQX-based NQS and independent PEPS calculations provided complementary evidence of spin-charge separation from spin pumping and real-time spin and charge dynamics. This application is distinct from the spin-polarized charge-pumping example documented here, but it demonstrates that the Hofstadter workflows extend to interacting spinful fermions and cross-method validation.

\subsection{Real-time tVMC}

The real-time calculation considers the density response to the removal of a local potential. An ACE reference state is first optimized for an open $8\times6$ Hofstadter lattice with $N=4$ spin-polarized particles, $V=2$, $\alpha=0.25$, and the pinning term $h_v n_0$ with $h_v=-4$. At $t=0$, the pinning term is removed and the optimized state initializes the ACE-PEFT evolution.

The ACE-PEFT variant applies Eq.~\eqref{eq:ace-peft}: the convolutional and site-wise nonlinear layers of the optimized reference state define a fixed, configuration-dependent neural basis, and only its complex final orbital map evolves. The calculation uses RK4 integration with time step $2\times10^{-3}$, $40960$ walkers, and the eigendecomposition solver with smooth cutoff $10^{-8}$.

\begin{figure}[t]
\centering
\includegraphics[width=0.58\linewidth]{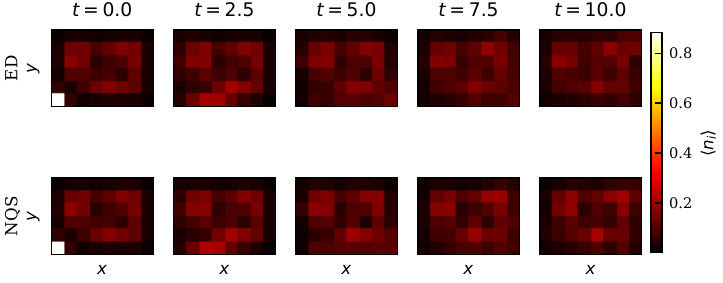}
\includegraphics[width=0.38\linewidth]{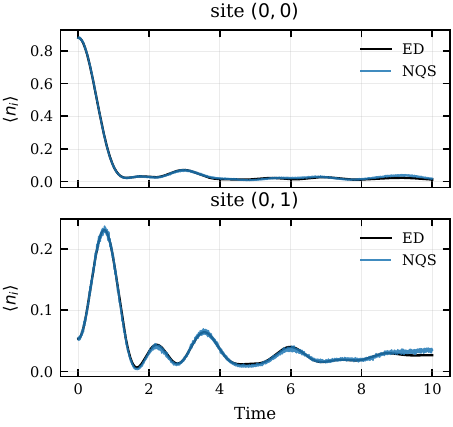}
\caption{Density dynamics following a local quench in the $8\times6$, four-particle Hofstadter model. Left: local-density snapshots from ED and tVMC at the nearest recorded times to $t=0,2.5,5,7.5,10$. Right: ED and tVMC density time series at sites $(0,0)$ and $(0,1)$. No additional temporal smoothing is applied. The stored tVMC observables are weighted averages over the four RK4 stages and are plotted at the corresponding step midpoints. Only the final affine orbital map of ACE-PEFT is evolved.}
\label{fig:tvmc}
\end{figure}

The density snapshots and time series in Fig.~\ref{fig:tvmc} track the redistribution of the initially pinned charge and reproduce the main features of the ED response.

\section{Conclusion}
\label{sec:conclusion}

We have introduced LaQX, a JAX implementation of neural quantum state methods for interacting lattice fermions and spins. The code combines Metropolis-Hastings sampling, several variational ansatz families, first-order and natural-gradient optimization, observables, stochastic projection, excited states, adiabatic continuation, and real-time evolution through a common set of functional interfaces. The doped Hubbard and frustrated $J_1$-$J_2$ Heisenberg model calculations connect the release to methods that have achieved high variational accuracy \cite{gu_hubbard_2025,gu_pareto_2026,fan_disentangling_2026}. The Hofstadter-Hubbard applications encompass many-body charge pumping \cite{hofstadter_pumping_2026} and NQS calculations of spin-charge separation \cite{gu_spincharge_2026}. The altermagnetic, excited-state, and dynamical examples test the same interfaces on one-body observables, low-energy states, and real-time evolution.

The library separates scientific components through explicit functional interfaces. Wavefunctions, physical operators, and update rules can therefore be extended through local source changes. This locality is useful for conventional development and for LLM-assisted programming, where unnecessary cross-file dependencies consume a finite context window. Source code, documentation, and the scripts used for the calculations in this paper are available at \url{https://github.com/bytedance/laqx}.

\section*{Acknowledgements}
We thank ByteDance Seed for inspiration and encouragement.

\bibliography{SciPost_Example_BiBTeX_File}

\end{document}